\def\be{\begin{equation}}
\def\ba{\begin{array}}
	\def\ea{\end{array}}

\documentclass[prl,showpacs,twocolumn,amsmath]{revtex4}
\usepackage{amsmath}
\allowdisplaybreaks
\usepackage{amsfonts}
\usepackage{mathrsfs}
\usepackage{amssymb}
\usepackage{pifont}
\usepackage{epsfig,subfigure,dsfont,amsthm,amsbsy,mathrsfs,amscd}
\usepackage{epstopdf}
\usepackage{color}
\usepackage{wrapfig}      
\usepackage{graphicx}
\usepackage[utf8]{inputenc}
\usepackage[T1]{fontenc}
\def\qed{\leavevmode\unskip\penalty9999 \hbox{}\nobreak\hfill
	\quad\hbox{\leavevmode  \hbox to.77778em{%
			\hfil\vrule   \vbox to.675em%
			{\hrule width.6em\vfil\hrule}\vrule\hfil}}
	\par\vskip3pt}
\usepackage{leftidx}

\newcommand{\adots}{\mathinner{\reflectbox{$\ddots$}}}
\begin{document}
	\title{\large\bf Quantum correlations of  tripartite mixed states in the black hole quantum atmosphere }

	\author{ Anqi Zhang$^{1}$, Yanze Zheng$^{1}$,  Xiaofen Huang$^{1, \dag}$ and Tinggui Zhang$^{1}$}
	\affiliation{ ${1}$ School of Mathematics and Statistics, Hainan Normal University, Haikou, 571158, China \\
		$^{\dag}$ huangxf1206@163.com }
	
	\bigskip
	\bigskip

	\begin{abstract}
    
We investigate quantum state texture, genuine multipartite entanglement, and tripartite nonlocality of a tripartite mixed state in the black hole quantum atmosphere. By introducing the Hartle-Hawking local temperature into the Bogoliubov coefficients, we characterize the influence of the local Hawking effect on both physically accessible and inaccessible reduced states. We find that the extrema of these three quantities all lie in the same near horizon region and shift outward with increasing local Hawking temperature, coinciding with the peak region of the local Hawking temperature and indicating that different aspects of tripartite quantum information are most sensitive to the local Hawking effect in the same atmospheric region. In contrast to genuine multipartite entanglement, tripartite nonlocality is more fragile and is suppressed under stronger local Hawking effects. These results provide a unified characterization of density matrix restructuring, entanglement redistribution, and nonlocality in tripartite mixed states affected by the black hole quantum atmosphere.

	\end{abstract}
	
	 \maketitle

\section{I. Introduction}

The combination of quantum information science and relativity has provided an important framework for studying quantum effects in noninertial frames and curved spacetimes \cite{AD,LB,PM,RB,RL}. In these relativistic settings, fundamental quantum resources, including entanglement \cite{entanglement1,entanglement2,entanglement3,entanglement4,genuine entanglement,concurrence1,concurrence2,Xiao2018, Tian2013, Wang2011a, Wang2010a, Wang2010c, Wang2010d, Wang2009}, entropic uncertainty relations \cite{EUR1,EUR2,EUR3,EUR4,EUR5}, quantum discord \cite{Wang2010c,discord1,discord2,discord3,discord4}, and nonlocality of Dirac fields \cite{Tian2013,Eur2022,Li2022,nonlocality1,nonlocality2,Svalue}, have been widely investigated. Black hole spacetime provides a particularly important gravitational scenario, where Hawking radiation links quantum field theory, gravity, and black hole thermodynamics. In the conventional description, Hawking radiation is usually associated with quantum excitations near the event horizon \cite{excitations1,excitations2,excitations3}. Therefore, understanding how the Hawking effect modifies and redistributes quantum correlations is essential for exploring quantum information in curved spacetime.

A central question in the study of Hawking radiation concerns the precise spatial location at which the emitted quanta are effectively generated. In the conventional view, Hawking radiation is usually attributed to quantum excitations located very close to the event horizon, namely \(\Delta r=r-r_H\ll r_H\), where \(r_H\) denotes the horizon radius \cite{Hawking1975,excitations3}. This near horizon interpretation has been challenged by the black hole quantum atmosphere proposal. According to Giddings' argument, the effective source region of Hawking radiation is not strictly confined to the event horizon, but extends to a finite region outside the black hole, with \(\Delta r=r_A-r_H\sim r_H\), where \(r_A\) denotes the effective radius of the quantum atmosphere \cite{Giddings2016}. This idea was originally motivated by estimates based on the Stefan-Boltzmann law and the wavelength of the emitted quanta, suggesting that the effective emission region extends well beyond the horizon scale \cite{Giddings2016}. It has since received further support from a variety of independent investigations, including studies of the stress energy tensor~\cite{Dey2017,Dey2019}, the gravitational Schwinger effect~\cite{Ong2020}, quantum correlations across the event horizon~\cite{Balbinot2022}, and the thermal behavior of the radiation~\cite{Kim2017,Eune2019}.

The quantum atmosphere framework therefore provides a natural setting for studying how the spatial origin of Hawking radiation affects quantum resources outside a black hole. The intrinsic quantum character of Hawking radiation makes the inspection of quantum correlations near the event horizon a suitable approach for probing possible signatures of the atmosphere. Previous studies have shown that nonlocal correlations can display a notable loss well outside the event horizon, which coincides with the peak of particle radiation in the atmosphere region \cite{Kaczmarek2024MIN}. Bipartite entanglement was also found to reach its extremum in the finite near horizon region \(1.43\lesssim r/r_h<1.5\), indicating a close relation between quantum resource extrema and the peak of the local Hawking radiation \cite{range}. More recently, multipartite coherence in the black hole quantum atmosphere has been investigated for GHZ-type states, where the \(l_1\)-norm coherence and the relative entropy of coherence exhibit atmosphere induced features, although these features become less visible as the number of parties increases \cite{Kaczmarek2025Coherence}. These works suggest that the quantum black hole atmosphere can leave detectable signatures in quantum resources. However, most existing studies still focus on bipartite correlations, pure multipartite states, or coherence measures, while the behavior of tripartite mixed state quantum information remains less explored.

In the present study, we explore the black hole quantum atmosphere from the perspective of tripartite mixed state quantum resources. We consider a tripartite mixed state comprising a generalized GHZ component and a maximally mixed noise component. This setting allows us to examine how white noise and the local Hawking effect jointly reshape multipartite quantum information. Specifically, we combine quantum state texture, genuine multipartite entanglement, and nonlocality to examine three complementary aspects of the tripartite mixed state. This framework reveals how tripartite mixed-state resources are redistributed in the quantum atmosphere and where the local Hawking effect becomes strongest.

The remainder of this paper is organized as follows. In Section II, we briefly review the vacuum structure of Dirac fields in Schwarzschild spacetime. In Section III, we construct a tripartite mixed state under the local Hawking effect and incorporate the Hartle-Hawking temperature to evaluate its quantum properties. We first analyze the quantum state texture of the physically accessible and inaccessible sectors, then examine the redistribution of genuine multipartite entanglement, and finally study tripartite nonlocality through the Svetlichny inequality, comparing its behavior with the other two quantities. The main findings are summarized in the concluding section.

\section{II. Vacuum structure of Dirac fields in Schwarzschild spacetime}	

We first recall the evolution process of the vacuum state and excited state of Dirac fields in the Schwarzschild black hole. In the natural units \(G=c=\hbar=k_{B}=1\), the metric of the Schwarzschild black hole can be written as
\begin{equation}
\begin{aligned}
ds^{2}=&-\left(1-\frac{2M}{r}\right)dt^{2}
+\left(1-\frac{2M}{r}\right)^{-1}dr^{2} \\
&+r^{2}\left(d\theta^{2}+\sin^{2}\theta d\phi^{2}\right),
\end{aligned}
\label{eq:schwarzschild_metric}
\end{equation}
where \(M\) denotes the mass of the black hole \cite{M}.
In the Schwarzschild spacetime, the Dirac equation for a massless fermionic field can be expressed as
\begin{equation}
\left[\gamma^{a}e^{\mu}_{a}\left(\partial_{\mu}+\Gamma_{\mu}\right)\right]\psi=0,
\label{eq:dirac_equation}
\end{equation}
where \(\gamma^{a}\) are the Dirac matrices, \(e^{\mu}_{a}\) is the inverse of the tetrad, and \(\Gamma_{\mu}\) denotes the spin connection coefficient. For the Schwarzschild geometry, Eq.~\eqref{eq:dirac_equation} takes the explicit form \cite{Dirac},
\begin{equation}
\begin{aligned}
&\frac{\gamma_{2}}{r}
\left(
\frac{\partial}{\partial\theta}
+\frac{\cot\theta}{2}
\right)\psi-\frac{\gamma_{0}}{\sqrt{1-\frac{2M}{r}}}\frac{\partial\psi}{\partial t} \\
&+\gamma_{1}\sqrt{1-\frac{2M}{r}}
\left[
\frac{\partial}{\partial r}
+\frac{1}{r}
+\frac{M}{2r(r-2M)}
\right]\psi \\
&+\frac{\gamma_{3}}{r\sin\theta}
\frac{\partial\psi}{\partial\phi}=0 .
\end{aligned}
\label{eq:dirac_schwarzschild}
\end{equation}
Solving Eq.~\eqref{eq:dirac_schwarzschild} near the event horizon, the positive fermionic frequency outgoing modes are obtained as
\begin{equation}
\psi^{I+}_{k}=\xi e^{-\mathrm{i}\omega u}, \qquad \psi^{II+}_{k}=\xi e^{\mathrm{i}\omega u},
\label{eq:schwarzschild_modes}
\end{equation}
where \(k\) represents the field mode, \(\omega\) is the monochromatic frequency of the Dirac field, \(\xi\) denotes the four component Dirac spinor and
\(u = t - r_{*}\), the tortoise coordinate is given by \(r_{*} = 2(M - D) \ln\left[\frac{r - 2M}{2(M - D)}\right] + r\). The solutions \(\psi^{I+}_{k}\) and \(\psi^{II+}_{k}\) correspond to the exterior and interior regions of the event horizon, respectively.
Using the Damour and Ruffini analytic continuation method, the positive energy Kruskal modes can be constructed from the Schwarzschild modes \cite{Damour}. The relation between the Kruskal modes and the Schwarzschild modes is given by
\begin{equation}
\begin{aligned}
\Phi^{+}_{k,I}\phantom{I}
&=e^{-2\pi M\omega}\Psi^{-}_{-k,II}
+e^{2\pi M\omega}\Psi^{+}_{k,I},\\
\Phi^{+}_{k,II}
&=e^{-2\pi M\omega}\Psi^{-}_{-k,I}
+e^{2\pi M\omega}\Psi^{+}_{k,II}.
\end{aligned}
\label{eq:kruskal_modes}
\end{equation}
Then the Dirac field can be quantized by the Schwarzschild modes and the Kruskal modes, respectively. In terms of the Kruskal modes, the Dirac field can be expanded as
\begin{equation}
\begin{aligned}
\psi
=&\int dk[2\cosh(4\pi M\omega_{i})]^{-\frac{1}{2}}
\big[
\hat{c}^{II}_{k}\Psi^{+}_{k,II}
+\hat{d}^{II\dagger}_{-k}\Psi^{-}_{-k,II} \\
&\quad
+\hat{c}^{I}_{k}\Psi^{+}_{k,I}
+\hat{d}^{I\dagger}_{-k}\Psi^{-}_{-k,I}
\big],
\end{aligned}
\label{eq:dirac_field_expansion}
\end{equation}
where \(\hat{c}_{k}\) and \(\hat{d}^{\dagger}_{-k}\) are the fermionic annihilation operator and the antifermionic creation operator, respectively.
According to the Bogoliubov transformation between the Schwarzschild and Kruskal operators, the Kruskal vacuum state and excited state in Schwarzschild spacetime are expressed as
\begin{equation}
|0\rangle_{k}
=
\frac{1}{\sqrt{e^{-\frac{\omega_{i}}{T}}+1}}
|0\rangle_{I}|0\rangle_{II}
+
\frac{1}{\sqrt{e^{\frac{\omega_{i}}{T}}+1}}
|1\rangle_{I}|1\rangle_{II},
\label{eq:kruskal_vacuum}
\end{equation}
and
\begin{equation}
|1\rangle_{k}=|1\rangle_{I}|0\rangle_{II},
\label{eq:kruskal_excited}
\end{equation}
where \(T\) denotes the Hawking temperature of the emitted radiation \cite{radiation}. For convenience, we denote the Bogoliubov coefficients by \(\mu=\frac{1}{\sqrt{e^{-\frac{\omega_i}{T}}+1}}\) and \(\nu=\frac{1}{\sqrt{e^{\frac{\omega_i}{T}}+1}}\), which satisfy \(\mu^2+\nu^2=1\). For simplicity, we take \(\omega_{i}=\omega=1\) in the following discussion. Here, \(|x\rangle_{I}\) and \(|x\rangle_{II}\) represent the fermionic modes outside the event horizon and the antifermionic modes inside the event horizon, respectively.

	\section{III. Tripartite Quantum Resources in the Black Hole Quantum Atmosphere}	
	\subsection{A. Quantum state texture affected by local Hawking radiation}
    
To characterize the quantum information of tripartite mixed states, we consider a class of states whose pure component is a generalized GHZ state and whose noisy component is the maximally mixed state. In this class of states, the GHZ component supplies genuine multipartite coherence, while the white noise component is controlled by the mixing parameter \(p\). It therefore provides a simple and tunable framework for analyzing how quantum information responds to coherence degradation and the degree of mixedness in a tripartite system.
The state is given by
\begin{equation}
\rho_{ABC}
=
p |\Psi_{\mathrm{GHZ}}\rangle\langle\Psi_{\mathrm{GHZ}}|
+\frac{1-p}{8} I_8,
\end{equation}
where
\begin{equation}
|\Psi_{\mathrm{GHZ}}\rangle
=
\alpha |000\rangle+\sqrt{1-\alpha^2}|111\rangle .
\end{equation}
Here, \(p\in[0,1]\) denotes the mixing parameter, which controls the relative contribution of the GHZ component, \(I_8\) is the identity matrix with size $8\times 8$, and \(\alpha\in[0,1]\) is the state parameter. The limits \(p=1\) and \(p=0\) correspond to the pure generalized GHZ state and the completely mixed state, respectively. By varying \(p\), one can systematically examine the effect of white noise on the quantum information of the tripartite state.

Quantum state texture has recently been proposed as a structural quantifier of quantum states. In a fixed computational basis, the density matrix may be viewed as a geometric landscape, and quantum texture measures its structural unevenness relative to the textureless state \(|f_1\rangle \equiv \frac{1}{\sqrt{d}}\sum_{i=1}^{d}|i\rangle\), with \(d\) being the dimension of the Hilbert space \(\mathcal{H}\). The quantum state texture is defined as \cite{Parisio} 
\begin{equation}
\mathcal{R}(\rho) = - \ln \langle f_1 | \rho | f_1 \rangle .
\label{eq:texture_def}
\end{equation}
This quantity has a direct geometric interpretation and can be measured through the projection probability onto the textureless state. Although related to coherence, quantum texture is not equivalent to conventional coherence measures and therefore provides an independent way to characterize quantum state structure.

In the tripartite mixed state considered above, the subsystems \(A\), \(B\), and \(C\) are associated with three observers, Alice, Bob, and Charlie respectively. Now we assume that Alice remains in the asymptotically flat region, whereas Bob and Charlie hover outside the event horizon of a Schwarzschild black hole at the same radial distance.

After applying the  transformation (\ref{eq:kruskal_vacuum}) and (\ref{eq:kruskal_excited}) to the modes of Bob and Charlie, we trace over the inaccessible regions \(B_{II}\) and \(C_{II}\), the physically accessible state \(\rho_{AB_I C_I}\) is obtained. This density matrix has an \(X\)-state structure in the standard computational basis, namely
\begin{equation}
\rho_{AB_I C_I}
=
\begin{pmatrix}
\rho_{11} & 0 & 0 & 0 & 0 & 0 & 0 & \rho_{18} \\
0 & \rho_{22} & 0 & 0 & 0 & 0 & 0 & 0 \\
0 & 0 & \rho_{33} & 0 & 0 & 0 & 0 & 0 \\
0 & 0 & 0 & \rho_{44} & 0 & 0 & 0 & 0 \\
0 & 0 & 0 & 0 & \rho_{55} & 0 & 0 & 0 \\
0 & 0 & 0 & 0 & 0 & \rho_{66} & 0 & 0 \\
0 & 0 & 0 & 0 & 0 & 0 & \rho_{77} & 0 \\
\rho_{81} & 0 & 0 & 0 & 0 & 0 & 0 & \rho_{88}
\end{pmatrix},
\end{equation}
where the matrix elements of the density operator $\rho_{AB_IC_I}$ are explicitly given by the following expressions:
\begin{equation*}
\begin{aligned}
\rho_{11} &= p \alpha^2 \mu^4 + \frac{1-p}{8}\mu^4, \\
\rho_{22} &= \rho_{33}
= p \alpha^2 \mu^2\nu^2
+ \frac{1-p}{8}\mu^2(1+\nu^2), \\
\rho_{44} &= p \alpha^2 \nu^4
+ \frac{1-p}{8}(1+\nu^2)^2, \\
\rho_{55} &= \frac{1-p}{8}\mu^4, \\
\rho_{66} &= \rho_{77}
= \frac{1-p}{8}\mu^2(1+\nu^2), \\
\rho_{88} &= p(1-\alpha^2)
+ \frac{1-p}{8}(1+\nu^2)^2, \\
\rho_{18} &= \rho_{81}
= p \alpha\sqrt{1-\alpha^2}\,\mu^2 .
\end{aligned}
\end{equation*}
By employing Eq.~\eqref{eq:texture_def}, we derive the quantum state texture of $\rho_{AB_IC_I}$:
\begin{equation}
\mathcal{R}(\rho_{AB_I C_I})
=
-\ln\!\left[
\frac{1+2p\alpha\sqrt{1-\alpha^2}\,\mu^2}{8}
\right].
\end{equation}
For comparison, we also evaluate the physically inaccessible states. Their explicit density matrices are listed in Appendix. These corresponding analyze expressions of  quantum state texture  are given by
\begin{equation}
\mathcal{R}(\rho_{AB_{II} C_{II}})
=
-\ln\!\left[
\frac{1+2p\alpha\sqrt{1-\alpha^2}\,\nu^2}{8}
\right].
\end{equation}

\begin{equation}
\mathcal{R}(\rho_{AB_I C_{II}})
=
\mathcal{R}(\rho_{AB_{II} C_I})
=
-\ln\!\left[
\frac{1+2p\alpha\sqrt{1-\alpha^2}\,\mu\nu}{8}
\right].
\end{equation}

To incorporate the effect of the black hole atmosphere, we replace the Hawking temperature $T$ in the Bogoliubov coefficients with the local temperature $T_{HH}$ in the Hartle-Hawking vacuum, which is given by
\begin{equation}
\begin{aligned}
T_{HH} &= T_H \sqrt{1-\frac{r_h}{r}}  \\
&\quad
\sqrt{
1+\frac{2r_h}{r}
+\left(\frac{r_h}{r}\right)^2
\left(9+4D_{HH}+36\ln\left(\frac{r_h}{r}\right)\right)
}.
\end{aligned}
\end{equation}
Here, \(T_H=\frac{1}{4\pi r_h}\) \cite{TH1} is the standard Hawking temperature, \(r_h\) denotes the event horizon radius, and \(r\) is the radial distance from the black hole center, with \(r>r_h\). The parameter \(D_{HH}\) is a constant associated with the stress tensor in the Hartle-Hawking vacuum. Since the Hartle-Hawking boundary conditions alone do not uniquely determine \(D_{HH}\), additional physical conditions are required. In order to avoid imaginary temperatures outside the event horizon and to keep the local temperature real in the physical region, we take \(D_{HH}\geq D_c\simeq 23.03\) \cite{range}.

From the above expression, one finds that the local temperature vanishes at the horizon, i.e., \(T_{HH}(r_h)=0\), and approaches the standard Hawking temperature in the asymptotic limit, \(T_{HH}\rightarrow T_H\) as \(r\rightarrow\infty\). Moreover, \(T_{HH}(r)\) reaches a maximum at a finite radial distance outside the event horizon. For \(D_{HH}\geq D_c\), the peak position is typically located in the interval \(1.43r_h\lesssim r_{\mathrm{peak}}<1.5r_h\) \cite{TH1}, where \(r_{\mathrm{peak}}\) denotes the position at which the local temperature is maximal. When \(D_{HH}\) is close to the critical value \(D_c\simeq23.03\), one has approximately \(r_{\mathrm{peak}}\simeq1.43r_h\). Thus, \(T_{HH}(r)\) characterizes how the local Hawking effect outside the event horizon varies with the radial position. In what follows, this temperature is used to determine the Bogoliubov coefficients, which describe how the black hole background modifies the field modes detected by observers close to the event horizon.

\begin{figure*}[t]
\centering

\begin{minipage}{0.32\textwidth}
\centering
\includegraphics[width=\textwidth]{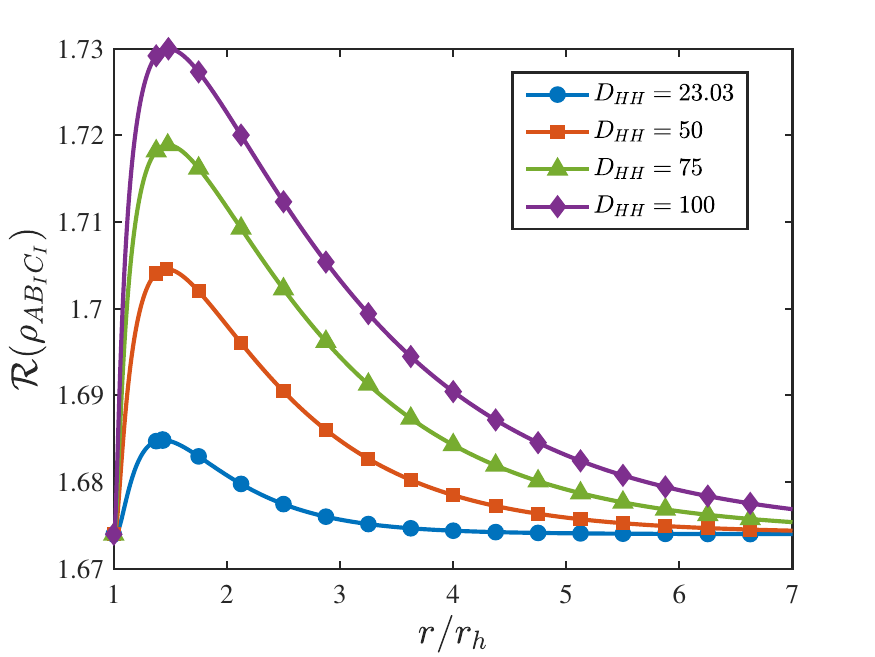}
\textbf{(a)}
\end{minipage}
\hfill
\begin{minipage}{0.32\textwidth}
\centering
\includegraphics[width=\textwidth]{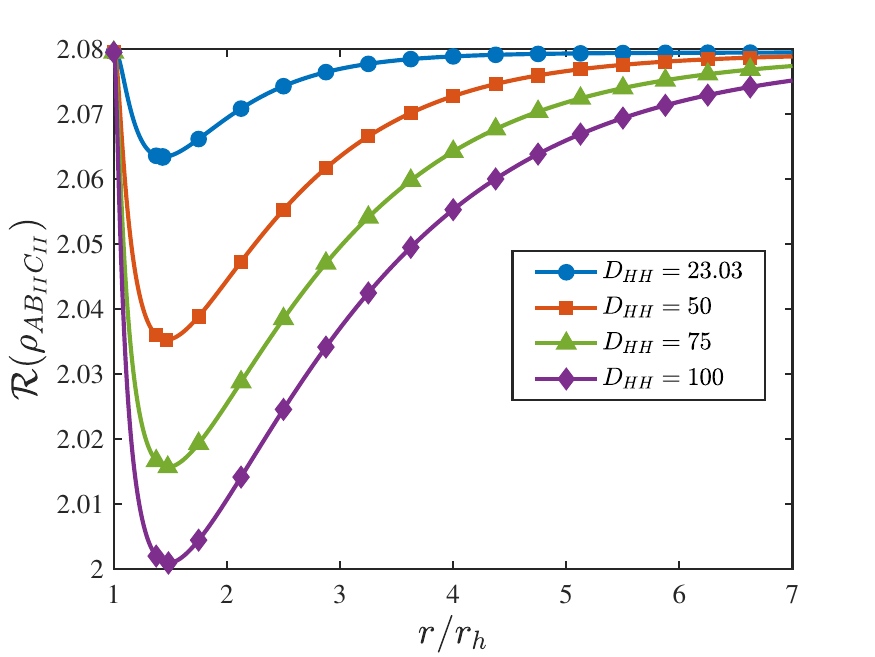}
\textbf{(b)}
\end{minipage}
\hfill
\begin{minipage}{0.32\textwidth}
\centering
\includegraphics[width=\textwidth]{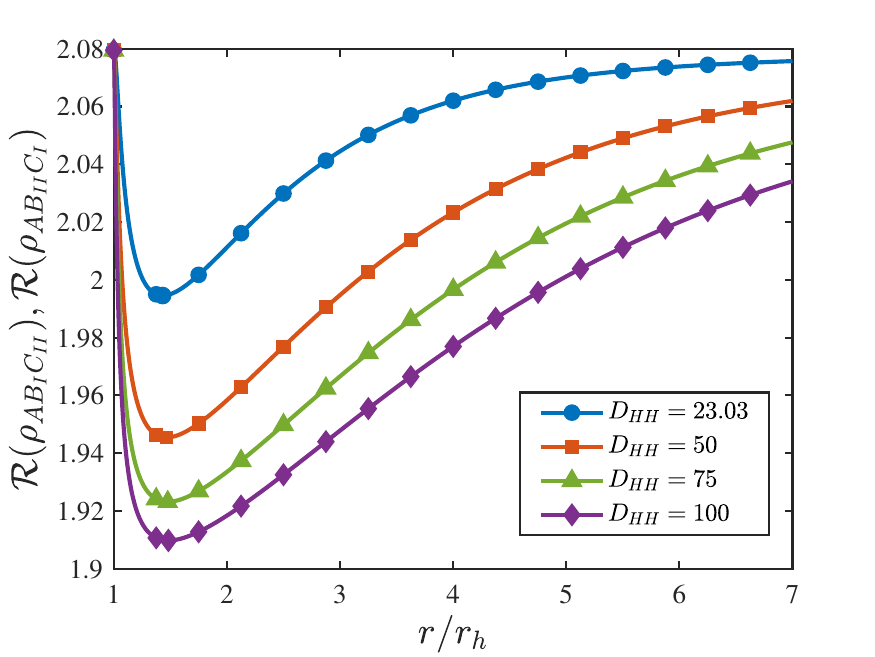}
\textbf{(c)}
\end{minipage}

\caption{
Quantum state textures versus the normalized distance \(r/r_h\) for different constants \(D_{HH}\).
Panel (a): \(\mathcal{R}(\rho_{AB_I C_I})\) versus \(r/r_h\);
Panel (b): \(\mathcal{R}(\rho_{AB_{II}C_{II}})\) versus \(r/r_h\);
and Panel (c): \(\mathcal{R}(\rho_{AB_I C_{II}})=\mathcal{R}(\rho_{AB_{II}C_I})\) versus \(r/r_h\).
The parameters are set as \(\alpha=\frac{\sqrt{2}}{2}\), \(p=\frac{1}{2}\).}
\label{fig:texture_r}
\end{figure*}

As shown in Fig. 1, the quantum state texture of both the physically accessible and inaccessible states exhibits clear local extrema with respect to \(r/r_h\). Specifically, the texture of \(\rho_{AB_I C_I}\) first increases and then decreases as the radial distance \(r/r_h\) grows, indicating that the quantum state structure in the accessible region is first enhanced near the horizon and then approaches a stable value. In contrast, the textures of the inaccessible states show the opposite trend, revealing that the local Hawking effect redistributes state structure between the two regions.

The position of the texture extremum depends on \(D_{HH}\).
For each fixed \(D_{HH}\), the texture extremum in all cases occurs at the same radial position, indicating that the radial response is governed by the same local Hawking temperature distribution.
As \(D_{HH}\) increases, the extremum shifts toward larger \(r/r_h\), but always remains within the range \(1.4322 \lesssim r/r_h < 1.5\). This interval is consistent with the extremal region of the local Hawking radiation in the quantum atmosphere, indicating that the quantum state texture of the tripartite mixed state tracks the radial peak of Hawking radiation. Thus, the texture extremum not only reflects the local restructuring of the quantum state but also serves as a quantum signature of near horizon Hawking radiation.

\subsection{B. Genuine multipartite entanglement under the local Hawking effect}            

 Entanglement is a central quantum resource, with applications in quantum communication, quantum computation, and quantum metrology \cite{AD, LB, RB, PM, RL}. Its behavior in relativistic settings has become an important topic in relativistic quantum information, especially for entanglement shared between inertial and noninertial observers \cite{RB, PM, RL, cgm}. Early work studied bipartite entanglement for bosonic and fermionic fields under acceleration \cite{RB, observer}, followed by extensions to cases where multiple observers are accelerated \cite{acceleration, two observers}. These studies showed that entanglement can be redistributed between accessible and inaccessible regions and that this redistribution depends on the initial state, field type, and chosen entanglement measure \cite{observer, acceleration, two observers, QJ19, EJ20, BM21, EI22, ME23, XM24, SN25}. For Dirac fields in curved spacetimes, the Hawking effect similarly modifies bipartite and multipartite entanglement \cite{EI22, ME23}.
 However, most existing studies are based on the standard Hawking temperature, while genuine multipartite entanglement in the black hole quantum atmosphere, where the local temperature depends on the radial distance, remains less explored.

We further investigate the influence of local Hawking radiation on genuine multipartite entanglement. Genuine multipartite entanglement characterizes quantum correlations that cannot be decomposed under any bipartite splitting of the multipartite system, and therefore offers an independent perspective on how the quantum atmosphere affects the quantum properties. To quantify this entanglement, we employ the genuine multipartite concurrence. If the density matrix of an $N$-qubit system has an $X$ form,
\begin{equation}
\rho_X=
\begin{pmatrix}
a_1      &        &        &        &        &        &        & z_1 \\
         & a_2    &        &        &        &        & z_2    &     \\
         &        & \ddots &        &        & \adots &        &     \\
         &        &        & a_n    & z_n    &        &        &     \\
         &        &        & z_n^*  & b_n    &        &        &     \\
         &        & \adots &        &        & \ddots &        &     \\
         & z_2^*  &        &        &        &        & b_2    &     \\
z_1^*    &        &        &        &        &        &        & b_1
\end{pmatrix}.
\label{eq:x_state_matrix}
\end{equation}
where \(n=2^{N-1}\), \(|z_i|\leq \sqrt{a_i b_i}\), and \(\sum_{i=1}^{n}(a_i+b_i)=1\), which ensure that \(\rho_X\) is positive and normalized. For this class of states, the genuine multipartite concurrence is given by \cite{genuine entanglement}
\begin{equation}
C_{\mathrm{GM}}(\rho_X)
=
2\max\left\{0,\ |z_i|-m_i\right\}, \quad i=1,\ldots,n,
\label{eq:concurrence}
\end{equation}
where
$
m_i = \sum_{j \neq i}^{n} \sqrt{a_j b_j}.
$
Since the physically accessible and inaccessible states exhibit an \(X\) structure, their genuine multipartite concurrence can be obtained directly by substituting the relevant matrix elements into Eq.~\eqref{eq:concurrence}. Based on this result, we further examine the dependence of genuine multipartite entanglement on \(r/r_h\), \(D_{HH}\), and the mixing parameter in both the physically accessible and inaccessible regions, thereby revealing the influence of local Hawking radiation on tripartite quantum correlations.

\begin{figure*}[t]
\centering

\begin{minipage}{0.32\textwidth}
\centering
\includegraphics[width=\textwidth]{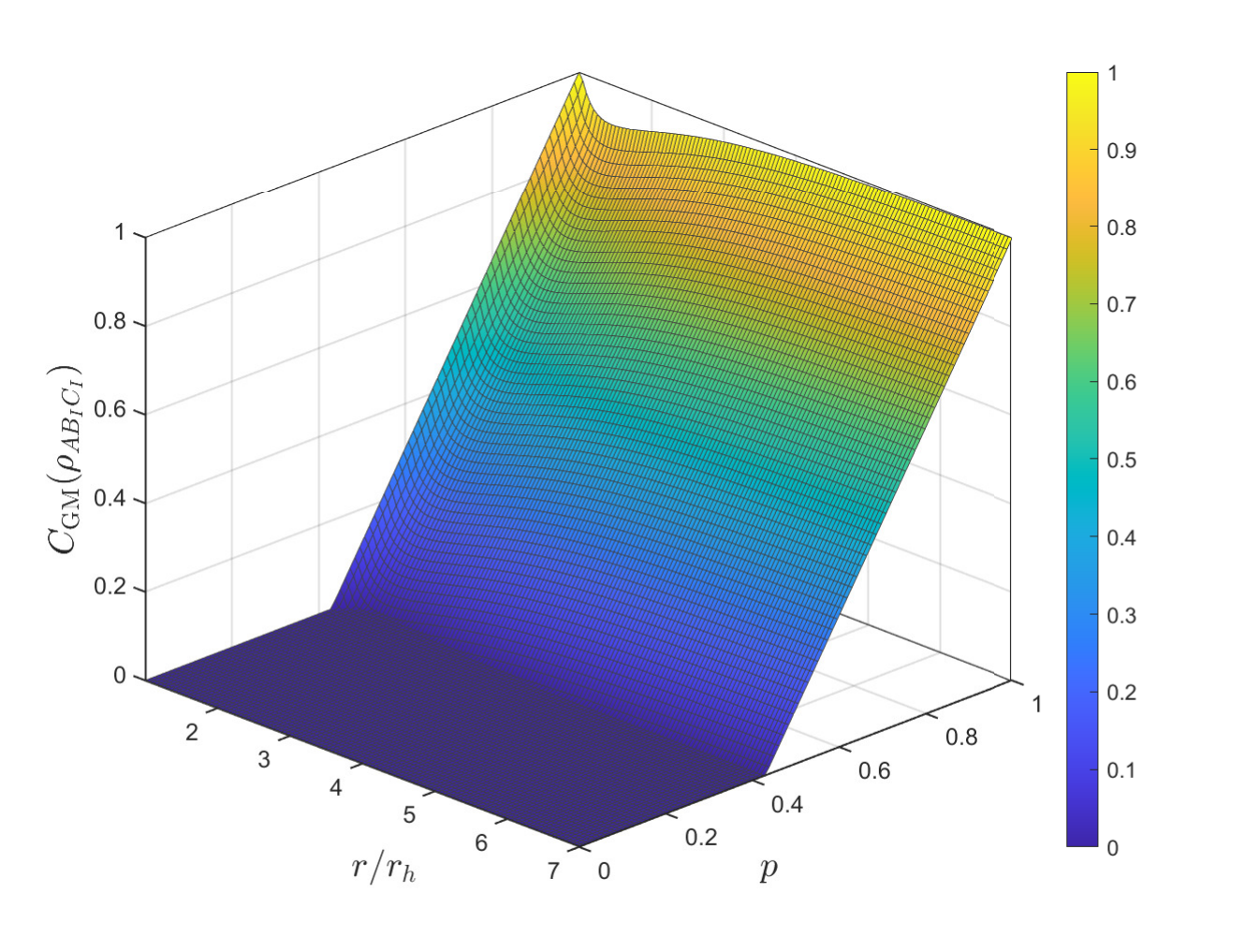}
\textbf{(a)}
\end{minipage}
\hfill
\begin{minipage}{0.32\textwidth}
\centering
\includegraphics[width=\textwidth]{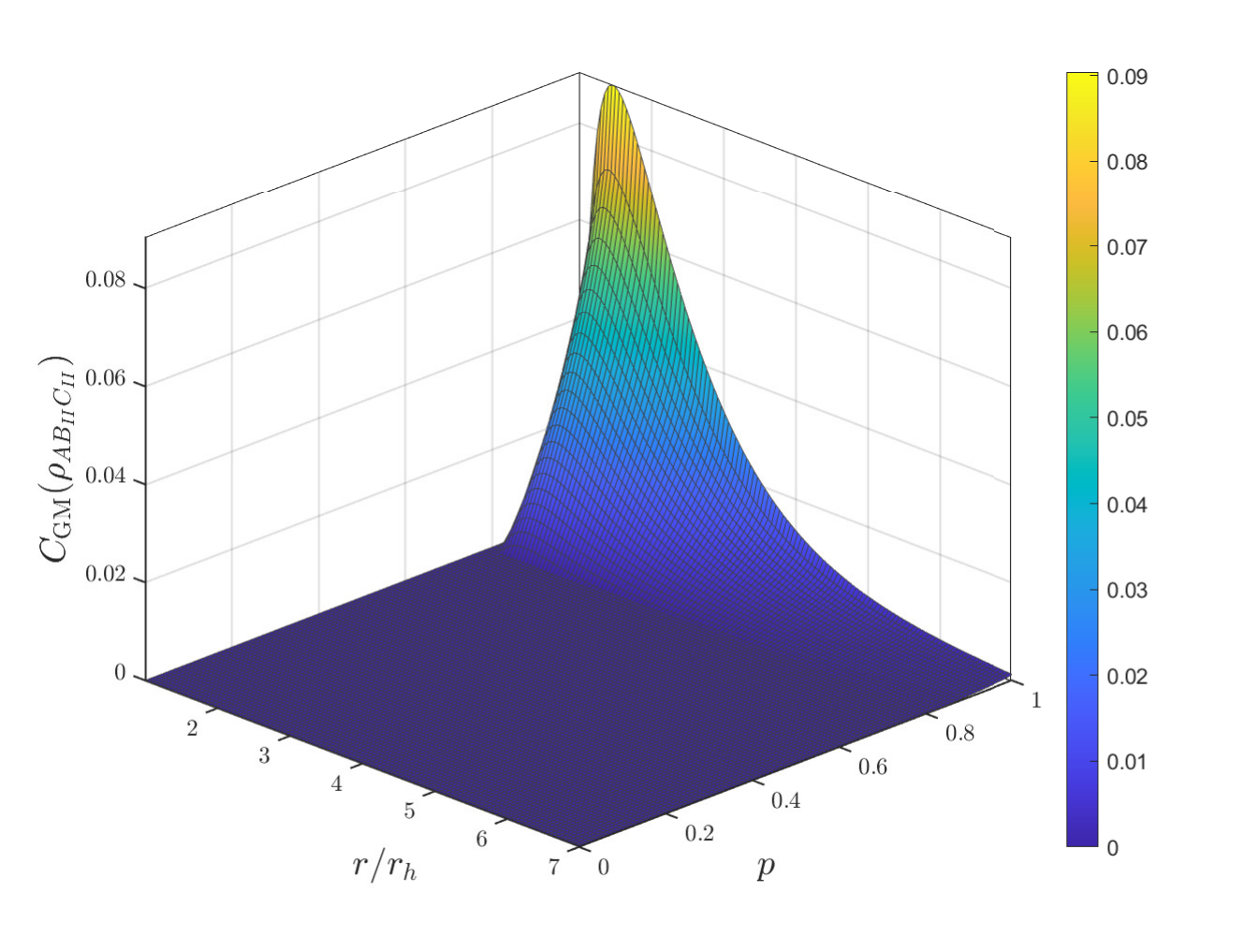}
\textbf{(b)}
\end{minipage}
\hfill
\begin{minipage}{0.32\textwidth}
\centering
\includegraphics[width=\textwidth]{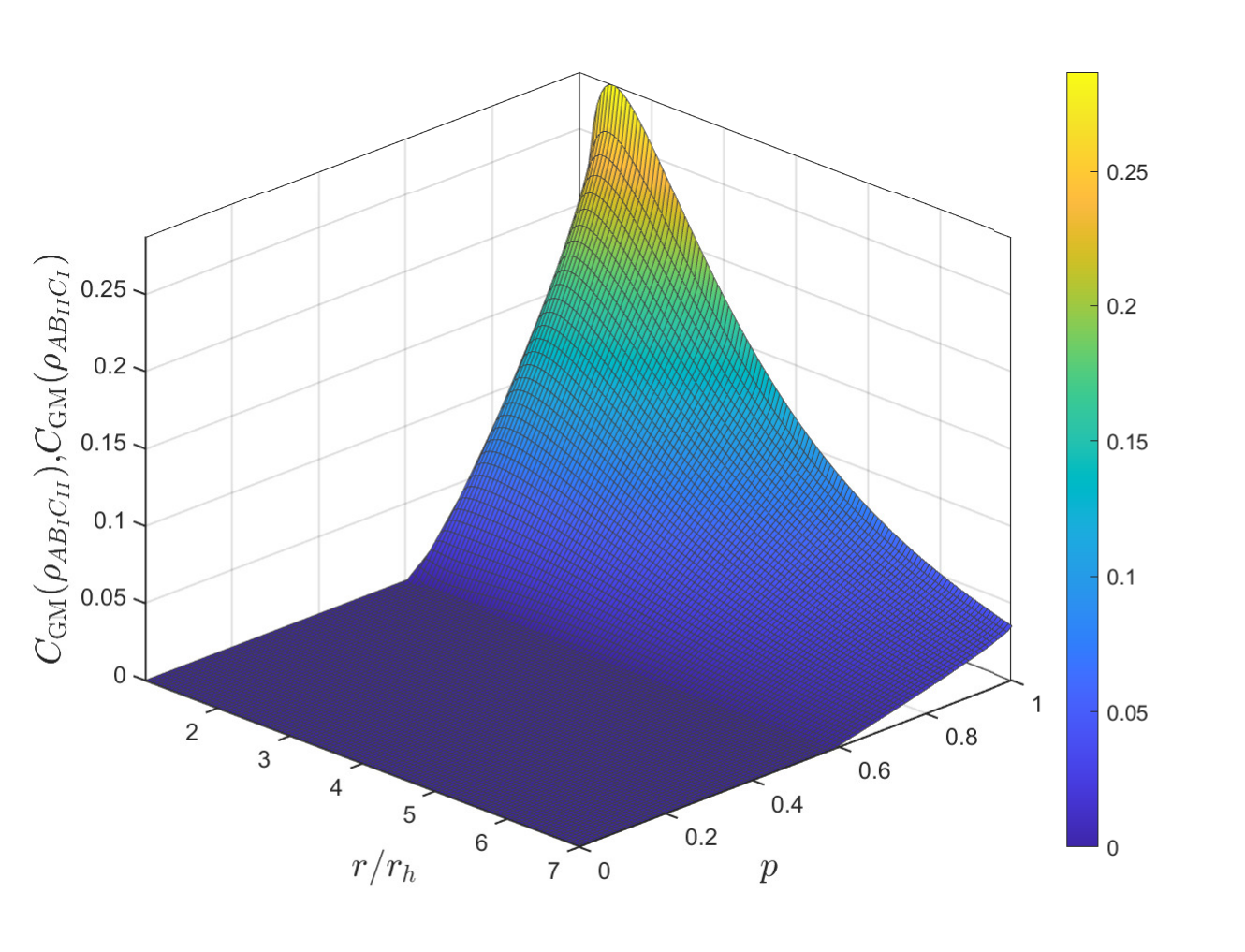}
\textbf{(c)}
\end{minipage}
\caption{Genuine multipartite concurrence \(C_{\mathrm{GM}}\) as a function of the Werner parameter \(p\) and the normalized radial distance \(r/r_h\) for the four reduced states, with \(D_{HH}=50\) and \(\alpha=\sqrt{2}/2\). 
}
\end{figure*}
Fig. 2 shows the genuine multipartite concurrence of the physically accessible and inaccessible reduces states as a function of the mixing parameter \(p\) and the normalized radial distance \(r/r_h\), with \(D_{HH}=50\) and \(\alpha = \frac{\sqrt{2}}{2}\). Overall, \(p\) plays a dominant role in determining the multipartite entanglement. For lower values of \(p\), the white noise component dominates the mixed state, and the entanglement is nearly absent in all four states. As $p$ increases, the GHZ component strengthens, and genuine multipartite entanglement gradually emerges and grows. The two states \(\rho_{AB_I C_{II}}\) and \(\rho_{AB_{II} C_I}\) are symmetric in \(r/r_h\) and therefore exhibit identical behavior.

The figure reveals that the physically accessible state exhibits the strongest and most widely distributed genuine multipartite entanglement, indicating that the dominant multipartite entanglement remains in the accessible region.
In contrast, the fully inaccessible state \( \rho_{AB_{II}C_{II}} \) exhibits much weaker entanglement, which is highly localized at large \(p\) and finite \(r/r_h\). The other two inaccessible states exhibit similar behavior.

Thus, Fig. 2 shows that the local Hawking effect induces a redistribution of genuine multipartite entanglement between the physically accessible and inaccessible states, while the accessible state remains the main carrier of multipartite entanglement.

\begin{figure*}[t]
\centering

\begin{minipage}{0.32\textwidth}
\centering
\includegraphics[width=\textwidth]{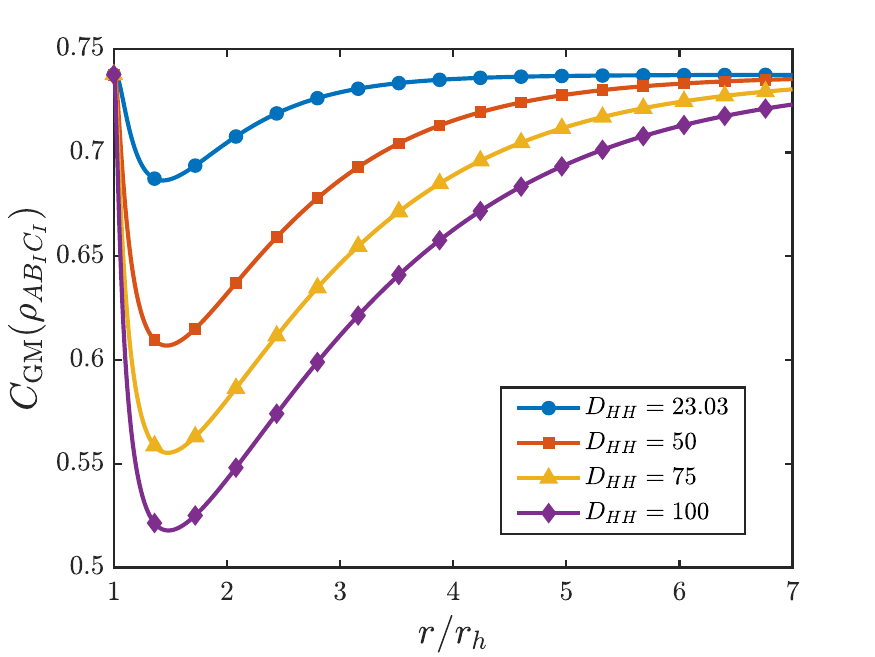}
\textbf{(a)}
\end{minipage}
\hfill
\begin{minipage}{0.32\textwidth}
\centering
\includegraphics[width=\textwidth]{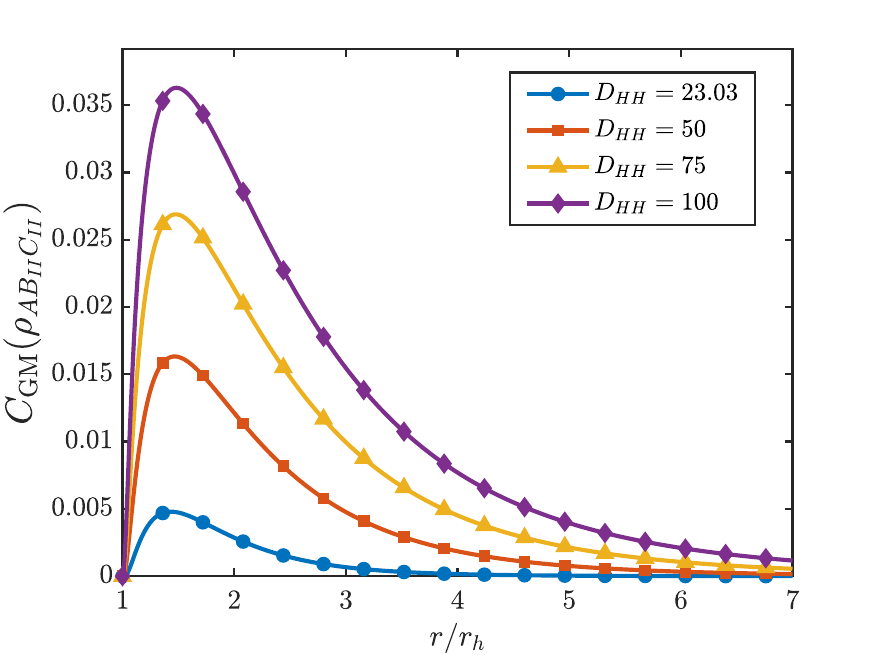}
\textbf{(b)}
\end{minipage}
\hfill
\begin{minipage}{0.32\textwidth}
\centering
\includegraphics[width=\textwidth]{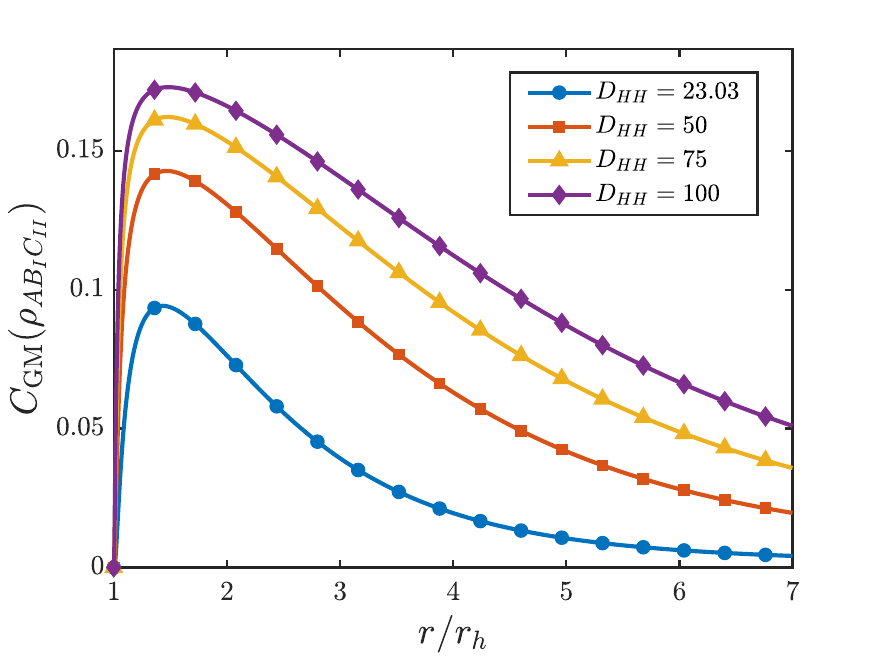}
\textbf{(c)}
\end{minipage}

\caption{
Genuine multipartite concurrence \(C_{\mathrm{GM}}\) versus the normalized radial distance \(r/r_h\) for different values of \(D_{HH}\). 
Panel (a): \(C_{\mathrm{GM}}(\rho_{AB_I C_I})\); 
Panel (b): \(C_{\mathrm{GM}}(\rho_{AB_{II}C_{II}})\); 
and Panel (c): \(C_{\mathrm{GM}}(\rho_{AB_I C_{II}})=C_{\mathrm{GM}}(\rho_{AB_{II}C_I})\). 
The parameters are fixed as $p=0.85$, $\alpha=\frac{\sqrt{2}}{2}$, and $D_{HH}=23.03, 50, 75, 100$.
}
\label{fig:texture_r}
\end{figure*}

To study the radial behavior of the local Hawking effect, we fix \(p = 0.85\) and \(\alpha = \frac{\sqrt{2}}{2}\) in Fig. 3 and plot \(C_{\mathrm{GM}}\) versus \(r/r_h\) for different \(D_{HH}\) in all cases. With this setting, the physically inaccessible reduced states exhibit genuine multipartite entanglement while preserving the white noise component in the mixed state, making it suitable for analyzing the local Hawking effect on entanglement distribution.

For the physically accessible state \(\rho_{AB_I C_I}\), the genuine multipartite concurrence \(C_{\mathrm{GM}}\) exhibits a local dip in a finite region outside the event horizon and then gradually recovers to a stable value as \(r/r_h\) increases.
By contrast, in the physically inaccessible states, \(C_{\mathrm{GM}}\) exhibits a localized peak: it is enhanced in a finite region near the event horizon and then decays to a small value as \(r/r_h\) increases. This indicates that the local Hawking effect does not simply destroy genuine multipartite entanglement but rather redistributes it between physically accessible and inaccessible states.

As shown in the Fig. 4, for fixed \(p = 0.85\) and \(r/r_h = 1.5\), \(C_{\mathrm{GM}}\) exhibits a non-monotonic behavior with respect to the state parameter \(\alpha\) and approaches zero as \(\alpha \to 0\) or \(\alpha \to 1\), indicating that genuine multipartite entanglement relies on the coherent superposition between the GHZ components. In the tripartite mixed state, the local Hawking effect modifies the optimal initial state structure for genuine multipartite entanglement, so that its maximum no longer necessarily corresponds to \(\alpha = \frac{\sqrt{2}}{2}\).

\begin{figure*}[t]
\centering

\begin{minipage}{0.32\textwidth}
\centering
\includegraphics[width=\textwidth]{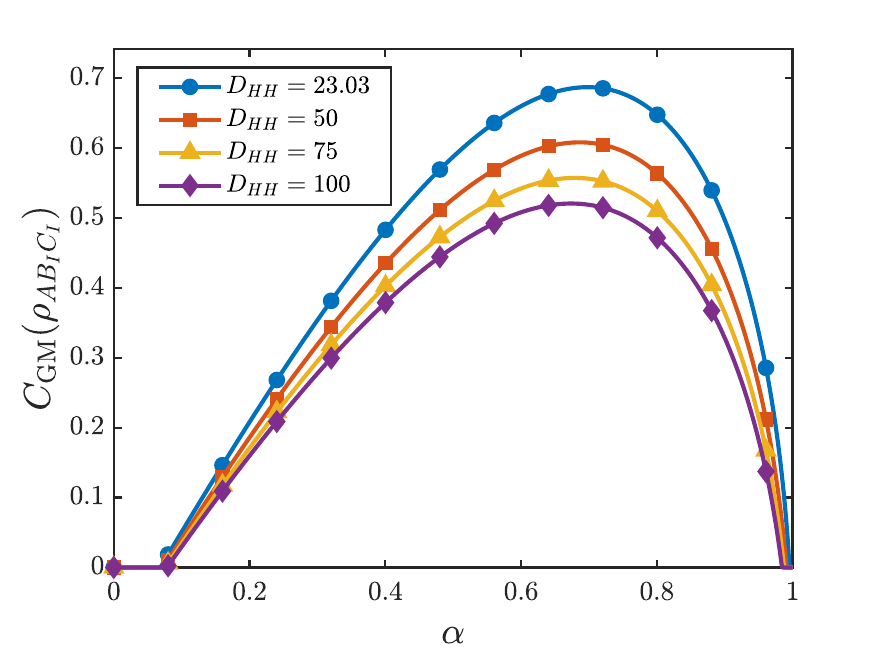}
\textbf{(a)}
\end{minipage}
\hfill
\begin{minipage}{0.32\textwidth}
\centering
\includegraphics[width=\textwidth]{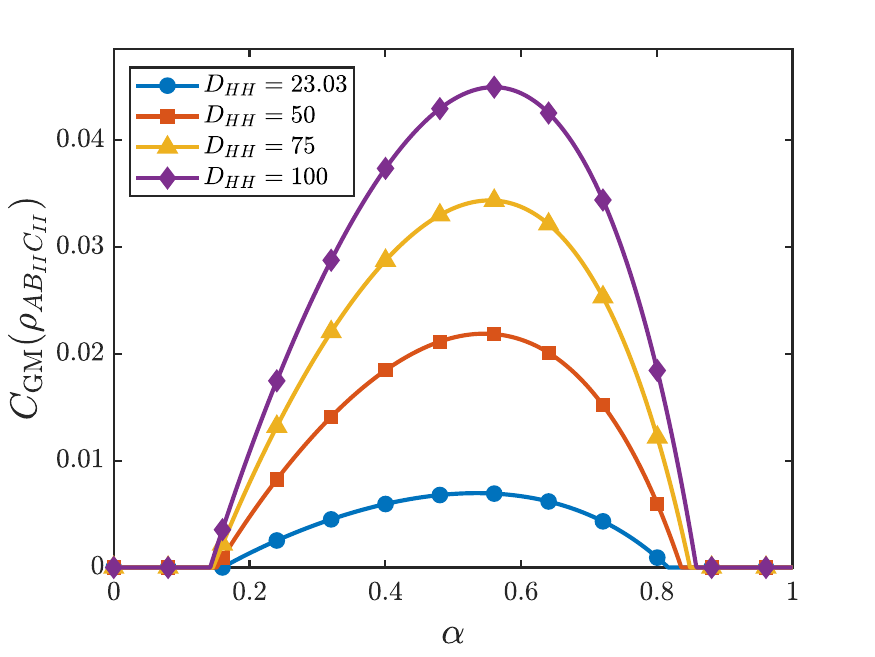}
\textbf{(b)}
\end{minipage}
\hfill
\begin{minipage}{0.32\textwidth}
\centering
\includegraphics[width=\textwidth]{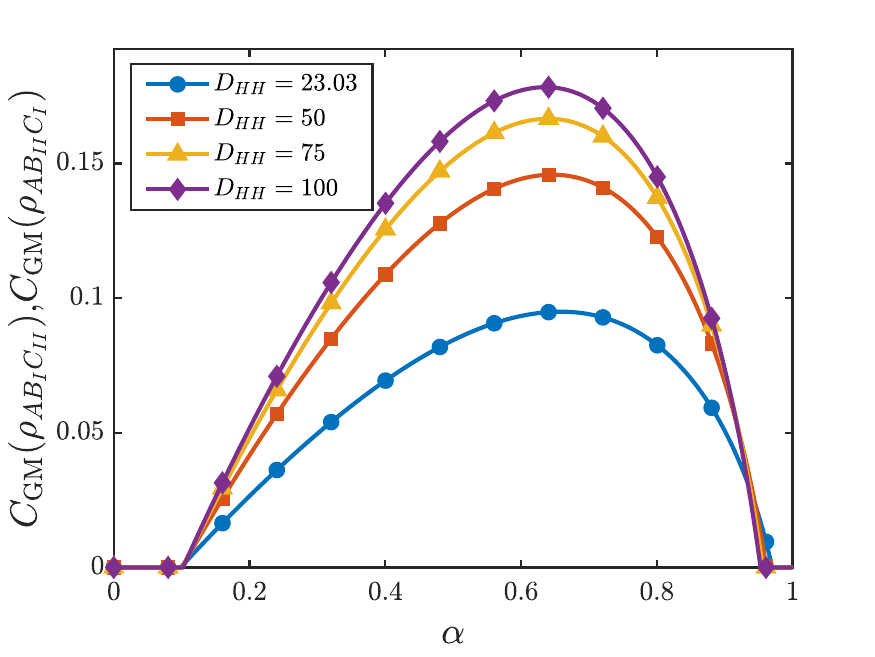}
\textbf{(c)}
\end{minipage}

\caption{
Genuine multipartite concurrence \(C_{\mathrm{GM}}\) versus the state parameter \(\alpha\) for different values of \(D_{HH}\), with \(r/r_h=1.5\) and \(p=0.85\). 
Panel (a): \(C_{\mathrm{GM}}(\rho_{AB_I C_I})\); 
Panel (b): \(C_{\mathrm{GM}}(\rho_{AB_{II}C_{II}})\); 
and Panel (c): \(C_{\mathrm{GM}}(\rho_{AB_I C_{II}})=C_{\mathrm{GM}}(\rho_{AB_{II}C_I})\). 
}
\label{fig:texture_r}
\end{figure*}

\subsection{C. Tripartite Nonlocality in the Black Hole Quantum Atmosphere}

Quantum nonlocality, originating from the Einstein-Podolsky-Rosen discussion on the incompleteness of quantum mechanics \cite{aiyinsitan}, is both a foundational concept and a useful resource for quantum information tasks \cite{tasks}, including device independent quantum key distribution \cite{key}, randomness expansion \cite{expansion}, and randomness amplification \cite{random}. For tripartite quantum systems, genuine tripartite nonlocality can be detected by the violation of the Svetlichny inequality, which excludes bipartite local hidden variable models. Let the local measurement operators be \(A_i=\vec{a}_i\cdot\vec{\sigma}\), \(B_i=\vec{b}_i\cdot\vec{\sigma}\), and \(C_i=\vec{c}_i\cdot\vec{\sigma}\) for \(i=0,1\), where \(\vec{a}_i\), \(\vec{b}_i\), and \(\vec{c}_i\) are real unit vectors and \(\vec{\sigma}=(\sigma_1,\sigma_2,\sigma_3)\) is the vector of Pauli matrices. The Svetlichny operator is
\begin{equation}
\begin{split}
S = & A_0(B_0 + B_1)C_0 + A_0(B_0 - B_1)C_1 \\
& + A_1(B_0 - B_1)C_0 - A_1(B_0 + B_1)C_1 .
\end{split}
\end{equation}
For a tripartite state \(\rho\) admitting a bilocal hidden variable model,
\begin{equation}
S(\rho) = |\operatorname{Tr}(S\rho)| \leq 4 .
\end{equation}
Thus, \(S(\rho)>4\) signals genuine tripartite nonlocality. In the black hole quantum atmosphere, the local Hawking effect can redistribute quantum correlations between physically accessible and inaccessible region. Therefore, it is important to determine whether such redistributed correlations remain strong enough to violate the Svetlichny inequality.

To quantify tripartite nonlocality, we consider a tripartite density matrix with \(X\) form in the computational basis \(\{|000\rangle,|001\rangle,\ldots,|111\rangle\}\),
\[
\rho_X =
\left(
\begin{array}{cccccccc}
d_1 & 0   & 0   & 0   & 0   & 0   & 0   & f_1 \\
0   & d_2 & 0   & 0   & 0   & 0   & f_2 & 0   \\
0   & 0   & d_3 & 0   & 0   & f_3 & 0   & 0   \\
0   & 0   & 0   & d_4 & f_4 & 0   & 0   & 0   \\
0   & 0   & 0   & f_4^{*} & e_4 & 0   & 0   & 0   \\
0   & 0   & f_3^{*} & 0   & 0   & e_3 & 0   & 0   \\
0   & f_2^{*} & 0   & 0   & 0   & 0   & e_2 & 0   \\
f_1^{*} & 0   & 0   & 0   & 0   & 0   & 0   & e_1
\end{array}
\right),
\]
with diagonal elements satisfying \(\sum_{i=1}^{4}(d_i+e_i)=1\). For this class of states, the Svetlichny value is given by
\begin{equation}
S(\rho_X) = \max \left\{ 8\sqrt{2}|f_i|, 4|\mathcal{N}| \right\},
\label{eq:Svalue}
\end{equation}
where \(\mathcal{N} = d_1 - d_2 - d_3 + d_4 - e_4 + e_3 + e_2 - e_1\) \cite{Svalue}.
The state exhibits genuine tripartite nonlocality when the Svetlichny inequality is violated, i.e., \(S(\rho_X) > 4\).

We now apply this criterion to the mixed state under the influence of the local Hawking effect. All of the physically accessible and inaccessible reduced states possess the X form, with their explicit density matrices given in Appendix. The Svetlichny values can be computed straightforwardly using Eq.~\eqref{eq:Svalue}.

\begin{figure*}[t]
\centering

\begin{minipage}{0.32\textwidth}
\centering
\includegraphics[width=\textwidth]{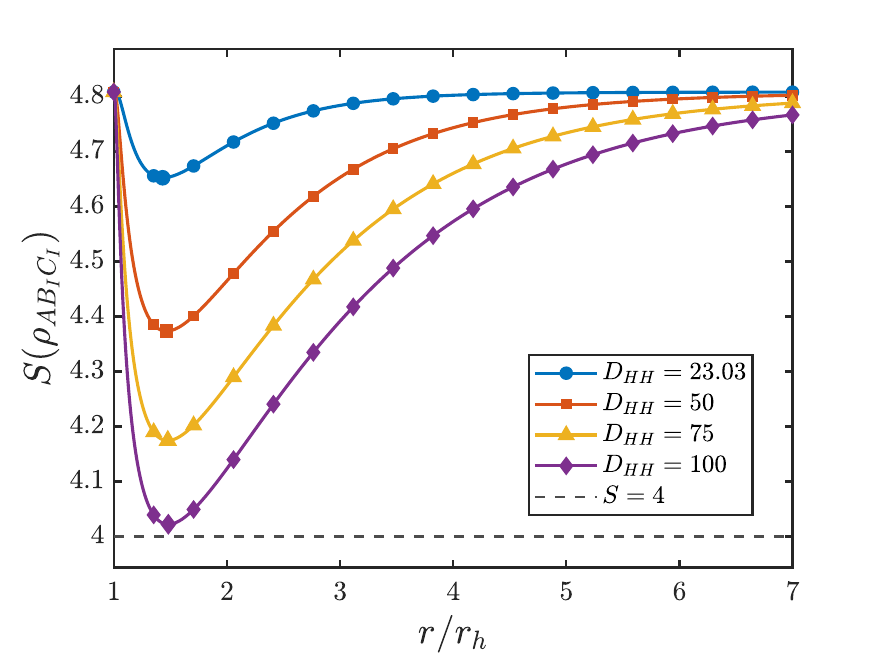}
\textbf{(a)}
\end{minipage}
\hfill
\begin{minipage}{0.32\textwidth}
\centering
\includegraphics[width=\textwidth]{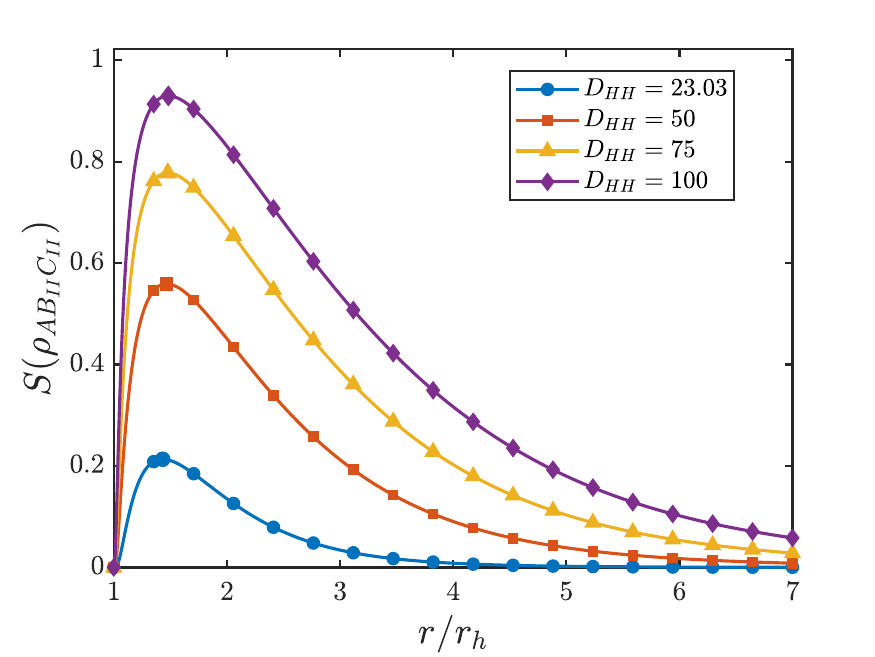}
\textbf{(b)}
\end{minipage}
\hfill
\begin{minipage}{0.32\textwidth}
\centering
\includegraphics[width=\textwidth]{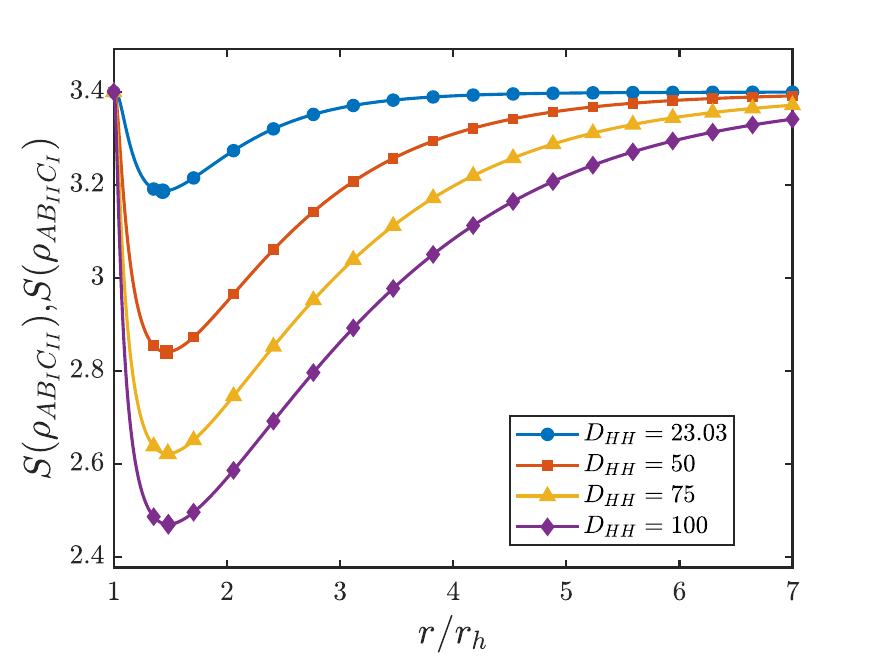}
\textbf{(c)}
\end{minipage}

\caption{Svetlichny value \(S\) versus the normalized radial distance \(r/r_h\) for different values of \(D_{HH}\).
Panel (a): \(S(\rho_{AB_I C_I})\);
Panel (b): \(S(\rho_{AB_{II}C_{II}})\);
and Panel (c): \(S(\rho_{AB_I C_{II}})=S(\rho_{AB_{II}C_I})\).
The parameters are fixed as \(p=0.85\), \(\alpha=\frac{\sqrt{2}}{2}\), and \(D_{HH}=23.03, 50, 75, 100\).
The black dashed line in Panel (a) denotes the Svetlichny nonlocality threshold \(S=4\).}
\label{fig:texture_r}
\end{figure*}

To further investigate the radial dependence of tripartite nonlocality and the effect of \(D_{HH}\), we now fix \(p = 0.85\) and plot \(S\) as a function of \(r/r_h\) for different values of \(D_{HH}\).

As shown in Fig. 5(a), the physically accessible state \(\rho_{AB_I C_I}\) exhibits a relatively large Svetlichny parameter \(S\) over the whole radial range. As \(r/r_h\) increases, \(S\) first decreases to a local minimum outside the event horizon and then rises again. For the relatively lower values of \(D_{HH}\) considered in this figure, \(S\) remains above the nonlocality threshold \(S=4\), indicating that the physically accessible state exhibits genuine tripartite nonlocality. Increasing \(D_{HH}\) suppresses the overall value of \(S\) and shifts the local minimum toward larger \(r/r_h\), indicating that the local Hawking effect weakens the tripartite nonlocality in the physically accessible state.

In contrast, the physically inaccessible states in Figs. 5(b) and 5(c) remain below the Svetlichny threshold \(S=4\) over the whole radial range. Although the detailed radial profiles depend on the specific reduced state, their Svetlichny values vary only weakly with \(r/r_h\), and no violation of the Svetlichny inequality is observed. This indicates that the local Hawking effect can affect the Svetlichny parameter in inaccessible regions, but the induced correlations do not exhibit genuine tripartite nonlocality. Tripartite nonlocality is therefore more fragile than genuine multipartite entanglement and is mainly retained in the physically accessible state.

\begin{table}[t]
\begin{ruledtabular}
\begin{tabular}{lccc}
\multicolumn{4}{c}{Extremal radial position \(r/r_h\)} \\
\hline
\(D_{HH}\) & \(\mathcal{R}\) & \(C_{\mathrm{GM}}\) & \(S\) \\
\hline
23.03 & 1.4322 & 1.4322 & 1.4322 \\
50    & 1.4667 & 1.4670 & 1.4670 \\
75    & 1.4772 & 1.4778 & 1.4778 \\
100   & 1.4832 & 1.4826 & 1.4826 \\
\end{tabular}
\end{ruledtabular}
\label{tab:extremal_positions}
\caption{Extremal radial positions of \(\mathcal{R}\), \(C_{\mathrm{GM}}\), and \(S\) for different \(D_{HH}\). The values correspond to the extrema of the physically accessible and inaccessible states.}
\end{table}

To better illustrate the radial behavior of quantum state texture, genuine multipartite entanglement, and genuine tripartite nonlocality, we summarize their extremal positions in Table I. As $D_{HH}$ increases from 23.03 to 100, the extrema shift outward from $r_{\mathrm{ext}}/r_h = 1.4322$ to approximately 1.48, while remaining within the finite near-horizon interval $1.4322 \lesssim r/r_h < 1.5$. 
Their extremal positions nearly coincide, suggesting that quantum state texture, genuine multipartite entanglement, and genuine tripartite nonlocality are all sensitive to the local Hawking effect in the same radial region. However, these quantities characterize distinct aspects: quantum state texture captures the structural redistribution of the density matrix, genuine multipartite entanglement reflects the distribution of multipartite correlations, and genuine tripartite nonlocality represents a stronger form of quantum correlation beyond standard entanglement. Thus, they serve as complementary signatures of the local Hawking effect, with their extrema consistently tracking the same near-horizon atmospheric region.

    \begin{figure}[htbp]
    	\centering
    	\includegraphics[width=0.4\textwidth]{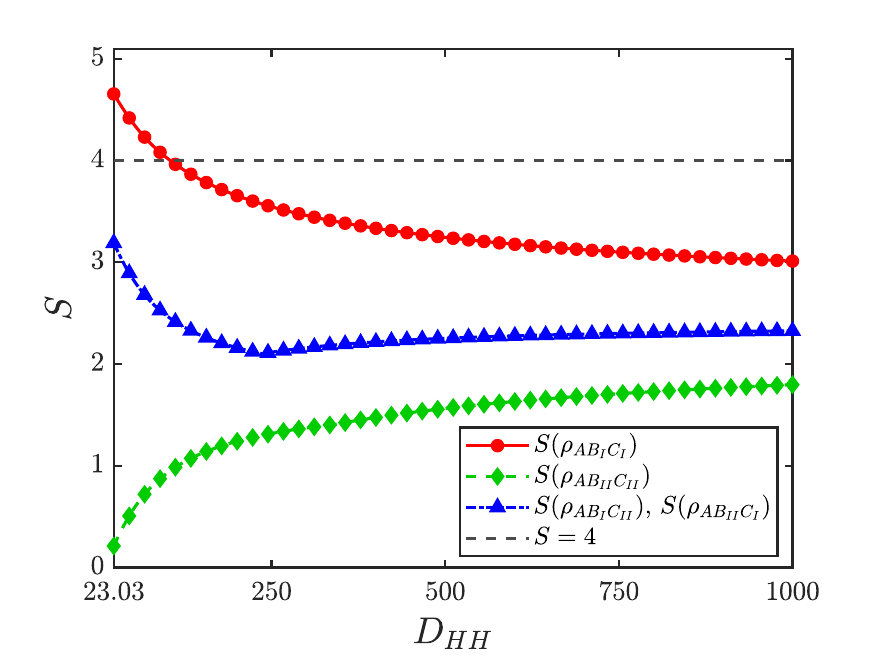}
    	\caption{
Plots of the Svetlichny value \(S\) as functions of \(D_{HH}\) at 
\(r/r_h=1.5\), \(p=0.85\), and \(\alpha=\frac{\sqrt{2}}{2}\). 
The curves correspond to \(\rho_{AB_I C_I}\), \(\rho_{AB_{II}C_{II}}\), and 
\(\rho_{AB_I C_{II}}\), respectively, and the dashed line denotes \(S=4\).
}
    	\label{Fig.3}
    \end{figure}

To further examine whether the tripartite nonlocality can survive under a stronger local Hawking effect, we study the Svetlichny parameter $S$ as a function of $D_{HH}$ for these reduced states. At fixed $r/r_h=1.5$, increasing $D_{HH}$ enhances the Hartle-Hawking local temperature and therefore strengthens the effective local Hawking effect. As shown in Fig. 6, $S(\rho_{AB_I C_I})$ decreases rapidly as $D_{HH}$ increases. It exceeds the nonlocality threshold $S=4$ only for small $D_{HH}$, and then drops below it as $D_{HH}$ increases. This shows that the genuine tripartite nonlocality of the physically accessible state is not universally preserved, but survives only when the local Hawking effect is relatively weak. As $D_{HH}$ increases further, the Svetlichny nonlocality is strongly suppressed and eventually disappears.

For the physically inaccessible states, the Svetlichny parameters also vary with $D_{HH}$, but remain strictly below the nonlocality threshold $S=4$ for all values of $D_{HH}$.
In particular, for $\rho_{AB_{II}C_{II}}$, $S$ increases with $D_{HH}$ but approaches an asymptotic value below the nonlocality threshold, so the enhancement of $S$ never induces genuine tripartite nonlocality.
For $\rho_{AB_I C_{II}}$ and $\rho_{AB_{II}C_I}$, $S$ exhibits nonmonotonic behavior: it first decreases and then slowly increases, eventually approaching a stable value. Overall, $D_{HH}$ influences the nonlocal correlations of all reduced states, with the most significant effects occurring at small $D_{HH}$. As $D_{HH}$ increases further, the variation weakens and gradually saturates. These results demonstrate that genuine tripartite nonlocality is more restrictive and fragile than genuine multipartite entanglement, and it cannot be generated solely by the redistribution of correlations into the inaccessible sectors.

\section{IV. Conclusions and discussions} 

In this work, we studied quantum state texture, genuine multipartite entanglement, and tripartite nonlocality of a tripartite mixed state in the black hole quantum atmosphere. By replacing the standard Hawking temperature with the Hartle-Hawking local temperature $T_{HH}(r)$, we incorporated the radial structure of the local Hawking effect into both physically accessible and inaccessible states. Our results show that these three quantities provide complementary signatures of the local Hawking effect: quantum state texture captures the restructure of density matrix, genuine multipartite entanglement reflects the redistribution of correlations between physically accessible and inaccessible sectors, and tripartite nonlocality characterizes the more restrictive condition of genuine nonlocal correlations. Although these quantities describe distinct physical aspects, their extremal positions all lie in the same region of near horizon  $1.4322 \lesssim r/r_h < 1.5$ and shift outward with increasing $D_{HH}$. This indicates that the quantum atmosphere provides a common radial window where tripartite quantum information is most sensitive to the local Hawking effect.

We further found that the mixing parameter $p$ controls the robustness of multipartite entanglement against white noise, while the state parameter $\alpha$ governs the initial GHZ structure. The local Hawking effect does not simply destroy genuine multipartite entanglement but redistributes it between accessible and inaccessible sectors. In contrast, tripartite nonlocality is more fragile: the physically accessible state violates the Svetlichny inequality only when the local Hawking effect is sufficiently weak, while the inaccessible states, despite retaining nonzero genuine multipartite entanglement, remain Svetlichny local. Therefore, these three quantities provide complementary levels of characterization for tripartite quantum resources in the quantum atmosphere of the black hole, from restructure of density matrix  to entanglement redistribution and strong nonclassical correlations.

\bigskip
\noindent{\bf Acknowledgments}
This work is supported by the Natural Science Foundation of Hainan Province under Grant No. 125RC744; the China Scholarship Council (CSC); the specific research fund of the Innovation Platform for Academicians of Hainan Province; the National Natural Science Foundation of China (NSFC) under Grant Nos. 12204137 and 12564048.

\bigskip
\textbf{Date Availability Statement}
All data generated or analyzed during this study are available from the corresponding author on reasonable request.

\bigskip
\textbf{Conflict of Interest Statement}
We declare that we have no conflict of interest.

\section*{Appendix: Density matrix of the tripartite reduced states}

We derive the density matrix $\rho_{AB_{II}C_{II}}$ by tracing out modes $B_I$ and $C_I$.
\begin{equation*}
\rho_{AB_{II}C_{II}}
=
\begin{pmatrix}
\rho_{11} & 0 & 0 & 0 & 0 & 0 & 0 & 0 \\
0 & \rho_{22} & 0 & 0 & 0 & 0 & 0 & 0 \\
0 & 0 & \rho_{33} & 0 & 0 & 0 & 0 & 0 \\
0 & 0 & 0 & \rho_{44} & \rho_{45} & 0 & 0 & 0 \\
0 & 0 & 0 & \rho_{54} & \rho_{55} & 0 & 0 & 0 \\
0 & 0 & 0 & 0 & 0 & \rho_{66} & 0 & 0 \\
0 & 0 & 0 & 0 & 0 & 0 & \rho_{77} & 0 \\
0 & 0 & 0 & 0 & 0 & 0 & 0 & \rho_{88}
\end{pmatrix}.
\end{equation*}
where the matrix elements of the density operator are explicitly given by the following expressions:
\begin{equation*}
\begin{aligned}
\rho_{11} &=
p\alpha^2\mu^4
+\frac{1-p}{8}(1+\mu^2)^2, \\
\rho_{22} &= \rho_{33}
=
p\alpha^2\mu^2\nu^2
+\frac{1-p}{8}(1+\mu^2)\nu^2, \\
\rho_{44} &=
p\alpha^2\nu^4
+\frac{1-p}{8}\nu^4, \\
\rho_{55} &=
p(1-\alpha^2)
+\frac{1-p}{8}(1+\mu^2)^2, \\
\rho_{66} &= \rho_{77}
=
\frac{1-p}{8}(1+\mu^2)\nu^2, \\
\rho_{88} &=
\frac{1-p}{8}\nu^4, \\
\rho_{45} &= p\alpha\sqrt{1-\alpha^2}\,\nu^2, \\
\rho_{54} &= \rho_{45}.
\end{aligned}
\end{equation*}
Finally, we present the explicit matrix forms of $\rho_{AB_{I}C_{II}}$ and $\rho_{AB_{II}C_{I}}$ , The corresponding density matrix $\rho_{AB_{II}C_{I}}$ takes the following form:
 \begin{equation*}
\rho_{AB_I C_{II}}
=
\begin{pmatrix}
\rho_{11} & 0 & 0 & 0 & 0 & 0 & 0 & 0 \\
0 & \rho_{22} & 0 & 0 & 0 & 0 & \rho_{27} & 0 \\
0 & 0 & \rho_{33} & 0 & 0 & 0 & 0 & 0 \\
0 & 0 & 0 & \rho_{44} & 0 & 0 & 0 & 0 \\
0 & 0 & 0 & 0 & \rho_{55} & 0 & 0 & 0 \\
0 & 0 & 0 & 0 & 0 & \rho_{66} & 0 & 0 \\
0 & \rho_{72} & 0 & 0 & 0 & 0 & \rho_{77} & 0 \\
0 & 0 & 0 & 0 & 0 & 0 & 0 & \rho_{88}
\end{pmatrix}.
\end{equation*}
where the matrix elements of the density operator are explicitly given by the following expressions:
\begin{align*}
\rho_{11} &=
p\alpha^2\mu^4
+\frac{1-p}{8}\mu^2(1+\mu^2), \\
\rho_{22} &=
p\alpha^2\mu^2\nu^2
+\frac{1-p}{8}\mu^2\nu^2, \\
\rho_{33} &=
p\alpha^2\mu^2\nu^2
+\frac{1-p}{8}(1+\nu^2)(1+\mu^2), \\
\rho_{44} &=
p\alpha^2\nu^4
+\frac{1-p}{8}(1+\nu^2)\nu^2, \\
\rho_{55} &=
\frac{1-p}{8}\mu^2(1+\mu^2), \\
\rho_{66} &=
\frac{1-p}{8}\mu^2\nu^2, \\
\rho_{77} &=
p(1-\alpha^2) + \frac{1-p}{8}(1+\nu^2)(1+\mu^2) \\
\rho_{88} &=
\frac{1-p}{8}(1+\nu^2)\nu^2, \\
\rho_{27} &=
p\alpha\sqrt{1-\alpha^2}\,\mu\nu, \\
\rho_{72} &= \rho_{27}.
\end{align*}
For completeness, the matrix form of $\rho_{AB_{II}C_{I}}$ takes the form,
\begin{equation*}
\rho_{AB_{II}C_I}
=
\begin{pmatrix}
\rho_{11} & 0 & 0 & 0 & 0 & 0 & 0 & 0 \\
0 & \rho_{22} & 0 & 0 & 0 & 0 & 0 & 0 \\
0 & 0 & \rho_{33} & 0 & 0 & \rho_{36} & 0 & 0 \\
0 & 0 & 0 & \rho_{44} & 0 & 0 & 0 & 0 \\
0 & 0 & 0 & 0 & \rho_{55} & 0 & 0 & 0 \\
0 & 0 & \rho_{63} & 0 & 0 & \rho_{66} & 0 & 0 \\
0 & 0 & 0 & 0 & 0 & 0 & \rho_{77} & 0 \\
0 & 0 & 0 & 0 & 0 & 0 & 0 & \rho_{88}
\end{pmatrix}.
\end{equation*}
where the matrix elements of the density operator are explicitly given by the following expressions:
\begin{equation*}
\begin{aligned}
\rho_{11} &=
p\alpha^2\mu^4
+\frac{1-p}{8}\mu^2(1+\mu^2), \\
\rho_{22} &=
p\alpha^2\mu^2\nu^2
+\frac{1-p}{8}(1+\mu^2)(1+\nu^2), \\
\rho_{33} &=
p\alpha^2\mu^2\nu^2
+\frac{1-p}{8}\mu^2\nu^2, \\
\rho_{44} &=
p\alpha^2\nu^4
+\frac{1-p}{8}\nu^2(1+\nu^2), \\
\rho_{55} &=
\frac{1-p}{8}\mu^2(1+\mu^2), \\
\rho_{66} &=
p(1-\alpha^2)
+\frac{1-p}{8}(1+\mu^2)(1+\nu^2), \\
\rho_{77} &=
\frac{1-p}{8}\mu^2\nu^2, \\
\rho_{88} &=
\frac{1-p}{8}\nu^2(1+\nu^2), \\
\rho_{36} &= p\alpha\sqrt{1-\alpha^2}\,\mu\nu, \\
\rho_{63} &= \rho_{36}.
\end{aligned}
\end{equation*}

\end{document}